\documentclass[aps,prl,reprint,superscriptaddress]{revtex4-2}
\usepackage{graphicx}
\usepackage{amsmath}
\usepackage{mathcomp}
\usepackage{textcomp}
\usepackage{siunitx}
\usepackage{physics}

\begin{document}


\title{Single photon randomness originating from the symmetry of dipole emission and the unpredictability of spontaneous emission}


\author{Michael Hoese}
\email[M.H. and M.K.K. contributed equally to this work.]{}
\affiliation{Institute for Quantum Optics, Ulm University, D-89081 Ulm, Germany}

\author{Michael K. Koch}
\email[M.H. and M.K.K. contributed equally to this work.]{}
\affiliation{Institute for Quantum Optics, Ulm University, D-89081 Ulm, Germany}
\affiliation{Center for Integrated Quantum Science and Technology (IQst), Ulm University, D-89081 Ulm, Germany}

\author{Felix Breuning}
\affiliation{Institute for Quantum Optics, Ulm University, D-89081 Ulm, Germany}
%


\author{Niklas Lettner}
\affiliation{Institute for Quantum Optics, Ulm University, D-89081 Ulm, Germany}

\author{Konstantin G. Fehler}
\affiliation{Institute for Quantum Optics, Ulm University, D-89081 Ulm, Germany}

\author{Alexander Kubanek}
\email[Corresponding author: ]{alexander.kubanek@uni-ulm.de}
\affiliation{Institute for Quantum Optics, Ulm University, D-89081 Ulm, Germany}
\affiliation{Center for Integrated Quantum Science and Technology (IQst), Ulm University, D-89081 Ulm, Germany}


\date{\today}

\begin{abstract}
Quantum random number generation is a key ingredient for quantum cryptography and fundamental quantum optics and could advance Monte-Carlo simulations and machine learning. An established generation scheme is based on single photons impinging on a beam splitter. Here, we experimentally demonstrate quantum random number generation solely based on the spontaneous emission process in combination with the symmetric emission profile of a dipole aligned orthogonal to the laboratory frame. The demonstration builds on defect centers in hexagonal boron nitride and benefits from the ability to manipulate and align the emission directionality. We prove the randomness in the correlated photon detection events making use of the NIST randomness test suite and show that the randomness remains for two independently emitting defect centers. 
The scheme can be extended to random number generation by coherent single photons with potential applications in solid-state based quantum communication at room temperature.
\end{abstract}


\maketitle


\section{Introduction}

The intrinsic randomness in quantum mechanical processes can be used for creating true, unpredictable random numbers via quantum random number generation (QRNG) \cite{Herrero2017}. Since QRNG devices start overcoming technical drawbacks, they become a promising replacement for conventional deterministic pseudo random number generators \cite{Wahl2011, Sanguinetti2014, Marangon2018}. QRNG is not only a key ingredient for fundamental experiments testing quantum mechanics \cite{Hensen2015} or quantum cryptography \cite{Peev2009, Sasaki2011}, but can also be exploited for Monte-Carlo simulations \cite{Ghersi2017} or machine learning \cite{Bird2019}. Photons impinging on a beam splitter provide a simple way to realize QRNG and has been used with LEDs to create true randomness \cite{Jennewein2000}. By employing single-photon emitters (SPE) such as single ions \cite{Pironio2010}, nitrogen-vacancy centers in diamond \cite{Chen2019} and quantum emitters in gallium nitride \cite{Luo2020} or hexagonal boron nitride (hBN) \cite{White2020} one can achieve a higher entropy per raw bit \cite{Oberreiter2015}.

In this work, we take a different approach exploiting the symmetry of the dipole emission pattern in order to realize QRNG. We make use of the fact that spontaneous emission is identical and random in all directions of the plane perpendicular to the dipole axis. Therefore, when observing the excited dipole from the left and from right in that plane, the spontaneous emission of a single photon happens with equal probability into one of the two observation channels. The randomness is given by the quantum mechanical nature of the spontaneous emission process, as illustrated in fig. \ref{Fig:TransmissionSetupSketch}(a), leading to high-quality RNGs and prohibiting any long-term correlations. 

It is desirable to utilize a SPE that is embedded in a solid-state matrix in order to realize compact devices among which 2D host materials play a special role \cite{Chakraborty2019}. Quantum emitters in hBN emerged as promising solid-state single photon source for photonic quantum technology \cite{Caldwell2019}. Technical overhead could be further reduced when the need for cryogenics is eliminated which requires detailed studies of electron-phonon interaction  \cite{Khatri2019, Grosso2020, Hoese2020}. Defect centers in hBN show, for example, almost unity quantum efficiency \cite{Boll2020},
giant strain-induced shift of the zero-phonon-line (ZPL) emission \cite{Li2020}
and, in particular, Fourier transform-limited (FTL) optical transitions under resonant excitation at cryogenic temperatures \cite{Dietrich2018} as well as room temperature (RT) \cite{Dietrich2020}. Remaining limitations, for example originating from on-going spectral diffusion and thermally generated phonon interactions under off-resonant excitation are currently investigated and continuously improved \cite{Akbari2021}. However, only special types of emitters have been identified to hold such coherent interactions under ambient conditions. These emitters are located in multi-layer hBN. The exact axial position and 3D orientation can, in general, be revealed with high accuracy \cite{Jha2020}. A study on the emission directionality suggests that the dipole is distorted out of the hBN plane such that the single photons are emitted parallel to the hBN layers \cite{Hoese2020}.


In this letter, we demonstrate single photon randomness without the use of a beamsplitter. We exploit the spontaneously emitted single photons from mechanically decoupled defect centers in a multi-layer hBN-flake. The emission is random in direction within the plane orthogonal to the dipole axis. Therefore, we align the dipole axis orthogonal to the optical axis of the setup by means of AFM-based nanomanipulation on the hBN host flake. We use the NIST randomness test suite to benchmark the randomness of the emitted photons and show that the random character persists for two SPE that emit single photons independently from each other. Hence, quantum emitters in hBN-flakes can be optimally fitted into optical setups for future application as QRNG devices.

\section{Experimental Setup}

\begin{figure}[]
\includegraphics[scale=1.]{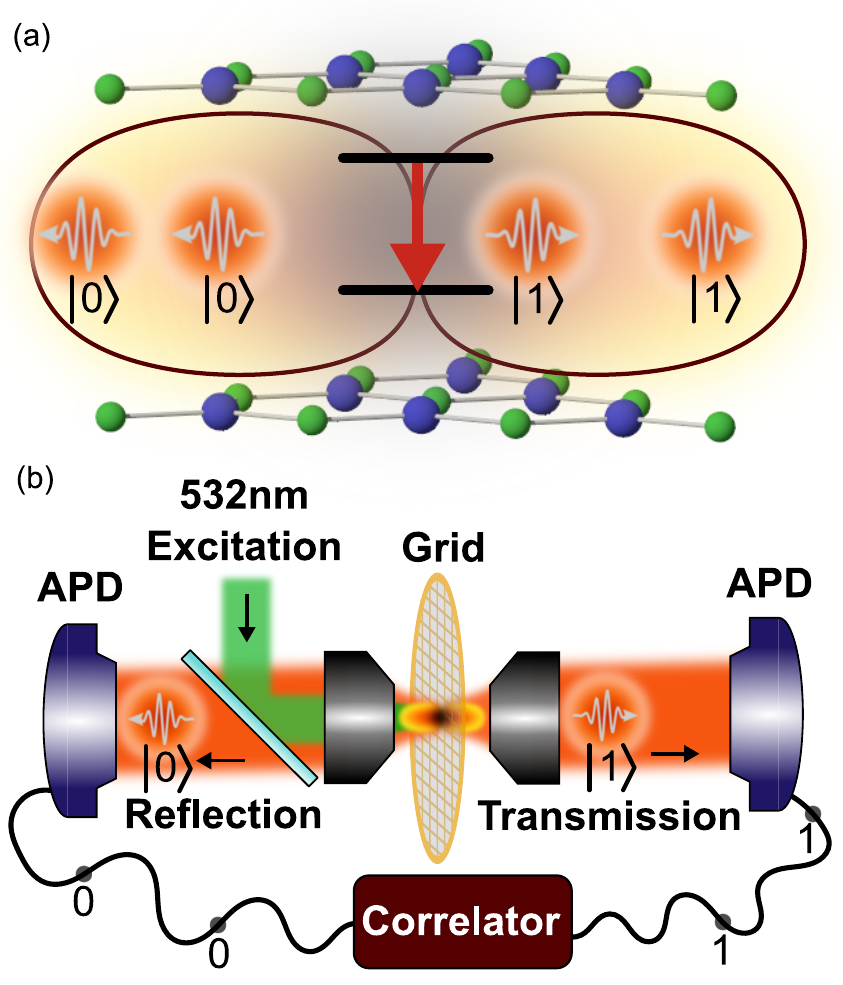}
\caption{\label{Fig:TransmissionSetupSketch} Experiment overview. 
(a) Experimental idea. The symmetry of the dipole emission pattern together with the intrinsic randomness of the spontaneous emission process generates randomness by encoding photons in forward and backward direction as 0 and 1, respectively. 
(b) Optical setup. The symmetric dipole emission pattern is accessed from two, opposite sites. The defect center in hBN is excited at a wavelength of 532 nm from the reflection side and spontaneously emits single photons into the reflection and transmission channel, where they are detected with avalanche photo diodes (APDs).}
\end{figure}

We begin by aligning the defect center in a hBN multi-layer flake with the optical setup. In order to align the dipole perpendicular to the optical axis with two detection ports from opposite sites, we use grids with low losses in transmission. A gold grid supports a \SI{20}{\nano\meter} thin perforated carbon foil with \SI{1.2}{\micro\meter} diameter holes \footnote{Quantifoil TEM Substrate R 1.2/1.3}. Ideally, the hBN-flakes are positioned within these holes. A quantum emitter in a perfectly-aligned flake emits photons randomly distributed towards the front and the back side with simultaneous optical access to both channels. We deposit hBN-flakes on the grid by slowly dragging it through an ethanol solution containing the hBN-flakes \footnote{Monolayer h-BN solution from 2D Semiconductors}. As shown in fig. \ref{Fig:TransmissionSetupSketch}(b), we extend a standard confocal setup with a transmission path. Therefore, we mount the grid between two objectives with high numerical aperture (NA = 0.9). On the front side, which we name reflection channel, the objective is mounted on a 3D scanning stage. In reflection the setup can be operated in standard confocal configuration with excitation and detection path and with access to an APD, a spectrometer for PL spectroscopy and a  Hanbury Brown and Twiss (HBT) setup for correlation measurements. The objective at the backside of the grid collects the light emitted in transmission into the second detection path, which we call transmission channel, again with access to APD, spectrometer and HBT setup. 


\section{AFM-based manipulation}

\begin{figure*}[]
\includegraphics[width=1.\textwidth]{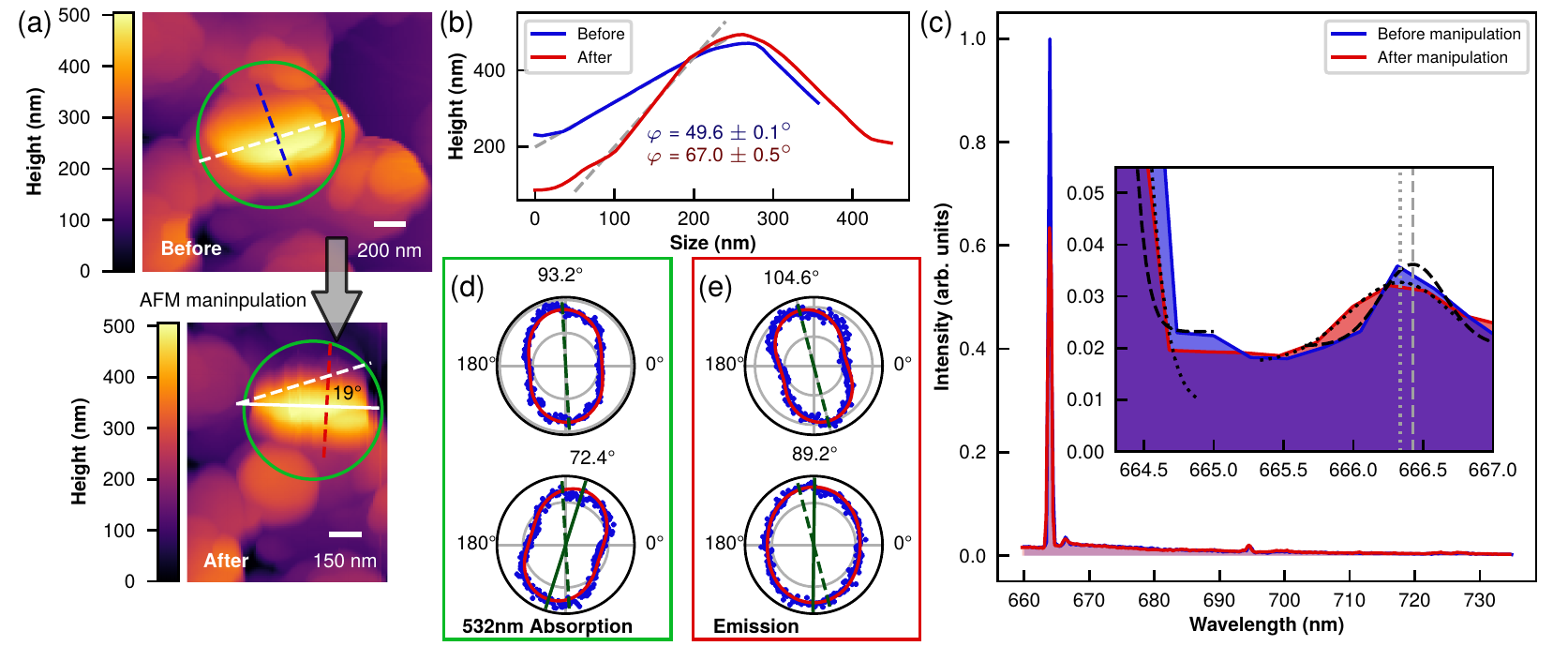}
\caption{\label{Fig:FlakeManipulationSpectrum}
Nanomanipulation of the hBN host flake. 
(a) The AFM image reveals the topology of the host flake (green circle) before (top) and after (bottom) AFM manipulation.  The dashed colored lines mark the cut for height profiles and white lines are used to determine the rotation angle.
(b) Height profiles before (blue) and after (red) AFM manipulation reveal an increased flake tilt.
(c) A comparison of the PL spectra before (blue) and after (red) manipulating the hBN host flake. The inset shows the energy gap between PSB and ZPL. Dashed and dotted lines indicate fits to the ZPL and first phonon mode before and after manipulation, with grey horizontal lines marking the positions of the first phonon mode.
(d)The \SI{532}{\nano\meter} absorption dipole orientation before (top) and after rotation (bottom, green lines mark dipole orientation before and after manipulation) of the hBN host flake.
(e) The emission dipole orientation changes accordingly when rotating the hBN host flake (before: top, after: bottom, green lines mark dipole orientation before and after manipulation)
}
\end{figure*}

Before we turn to QRNG, we proof the capability to align the emission intensity and polarization pattern of the defect center with respect to the optical setup without changing the optical properties. In particular persistence of the mechanical isolation of the defect center against nanomanipulation is important since it enables coherent photon emission.
We first demonstrate manipulation of the relative emission directionality of the hBN-flake with respect to the laboratory frame. For proof-of-principle demonstration, we spin-coat the hBN-flakes onto a sapphire substrate. We precharacterize the spectral properties of the emitter and determine the topology of its host flakes with an AFM. The mechanical isolation of the emitter is evidenced by the energy gap between ZPL and phonon sideband (PSB) \cite{Hoese2020}. To rotate the emission direction, we push the flake upright with the AFM in contact mode (fig. \ref{Fig:FlakeManipulationSpectrum}(a)) increasing the tilt angle from $49.6 \pm 0.1\si{\degree}$ to $67.0 \pm 0.5\si{\degree}$ (fig. \ref{Fig:FlakeManipulationSpectrum}(b)). The PL spectra of the emitter before and after nanomanipulation reveals a decreased intensity by the factor 0.6 (fig. \ref{Fig:FlakeManipulationSpectrum}(c)). Notably, the gap size between ZPL and PSB remains unchanged, thus indicating that the mechanical isolation of the emitter is unaffected by nanomanipulation.

 
Moreover, we rotate the flake by $19\si{\degree}$, as illustrated in fig. \ref{Fig:FlakeManipulationSpectrum}(a) and accordingly the polarization of the absorption and emission dipole. Polarization measurements before and after manipulation yield a rotation of the \SI{532}{\nano\meter} absorption dipole by $20.8 \pm 1.2\si{\degree}$ (see fig. \ref{Fig:FlakeManipulationSpectrum}(d)) and of the emission dipole by $15.3 \pm 1.4\si{\degree}$ (see fig. \ref{Fig:FlakeManipulationSpectrum}(e)). 
Controlled rotation of the hBN-flake enables to optimize and align the absorption and emission polarization to the optical axis of the experimental setup.

\section{Emitter characterization on the grid}

\begin{figure*}[]
\includegraphics[scale=1.0]{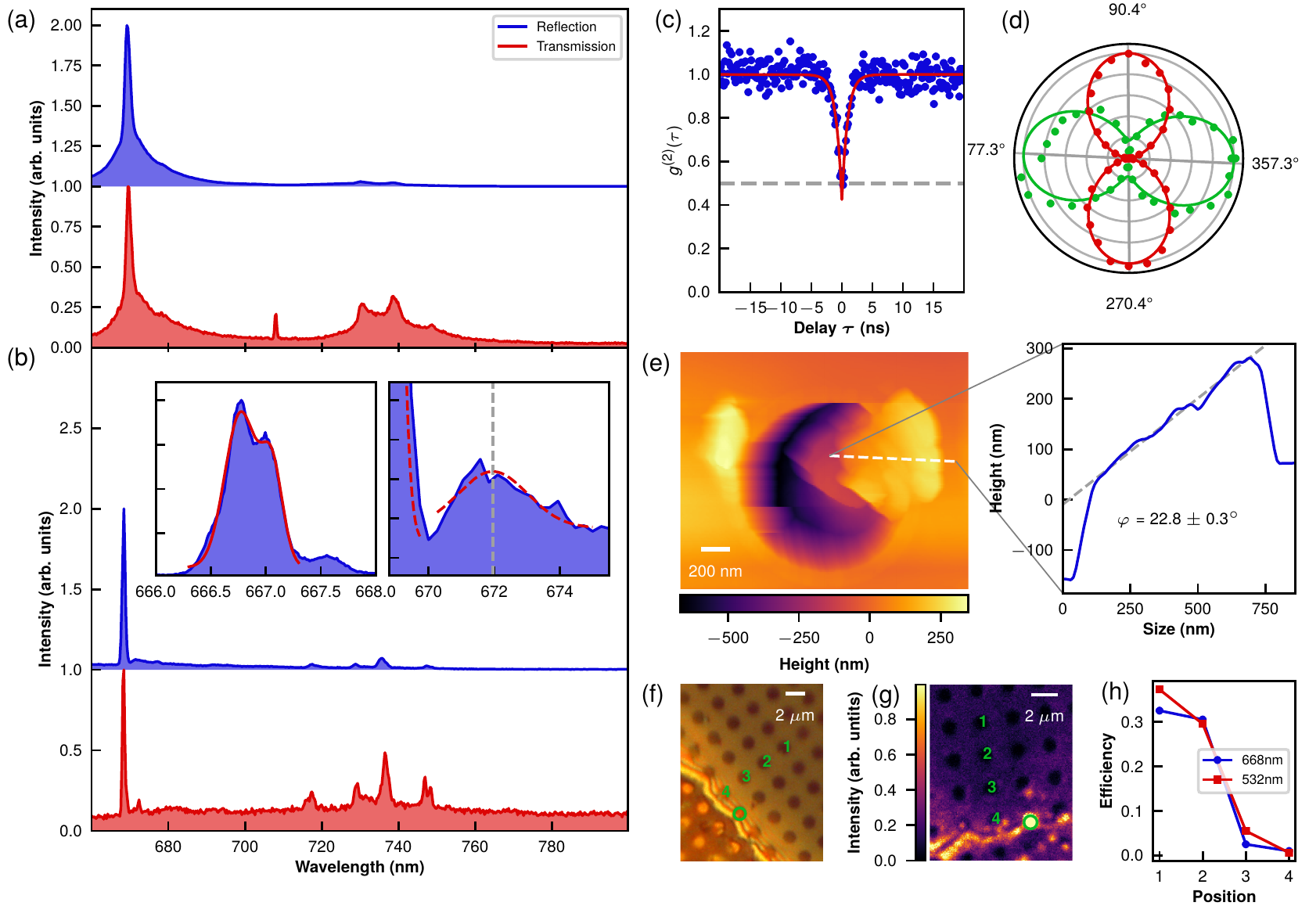}
\caption{\label{Fig:TEMGridOverview}
Spectroscopic characterization of the defect center in hBN on the transmission grid. (a) The room temperature PL spectra are compared for the reflection (blue) and transmission (red) channel.
(b) The PL spectra in reflection (blue) and transmission (red) are compared at \SI{5}{\kelvin}. The inset shows a high-resolution PL spectrum of the ZPL (left panel) and the gap between PSB and ZPL (right panel) with Gaussian fits (red). 
(c) A HBT correlation measurement at \SI{5}{\kelvin} yields $g^{\left(2\right)}\left(\tau\right) = 0.42 \pm 0.03$.
(d) The emitter yields distinct \SI{532}{\nano\meter} absorption (green) and emission dipoles (red).
(e) The topology of the hBN host flake discloses the tilt into one hole of the carbon foil of the grid. To the right, we plot the height profile along the dashed white line.
(f) The microscope image shows the area around the hBN host flake of the emitter (green circle). A series of holes of the grid is marked for comparison.
(g) The confocal image of the area around the hBN host flake (green circle) complements the microscope image.
(h) The transmission efficiency of green and red light through the grid at positions as labeled before.
}
\end{figure*}

In order to prepare the experiment, we place a defect center in a hBN-flake onto the transmission grid such that its dipole is as good as possible aligned orthogonal to the optical axis of the setup.
A comparison of the PL spectra in reflection and transmission yields the same distinct ZPL at \SI{667}{\nano\meter} with low background fluorescence, both at room temperature (RT) and at \SI{5}{\kelvin} (see fig. \ref{Fig:TEMGridOverview}(a) and (b)). Our setup works from RT to liquid helium temperature (\SI{5}{\kelvin}), thus allowing for in-depth spectral characterization. Moreover, our QRNG scheme in principle works at cryogenic temperatures, which could be useful for improved spectral properties and higher efficiency. The smaller peaks between \SI{710}{\nano\meter} and \SI{750}{\nano\meter}, which are pronounced in transmission, could originate either from optical phonon modes or from other defect centers and are spectrally filtered out in the following. Otherwise, the investigated emitter shows ideal spectral properties with a sharp, intense ZPL at \SI{667}{\nano\meter} and low background signal. However, the high-resolution PL spectrum of the ZPL reveals the double defect nature of the emitters \cite{Bommer2019}, as the ZPL contains two lines separated by \SI{0.3}{\nano\meter} (see inset in fig. \ref{Fig:TEMGridOverview}(b)). The second optical transition arises from a second emitter and lies within the spectral window of the detection paths, although less bright in intensity. We claim, that the emitted photon stream originating from two independent single photon sources do not show any memory or long-time correlations. Therefore, QRNG is still possible with increased rates by up to a factor of two. Accordingly, the corresponding correlation measurements yield $g^{\left(2\right)}\left(0\right) = 0.42 \pm 0.03$ at \SI{5}{\kelvin} as depicted in fig. \ref{Fig:TEMGridOverview}(c). 
Again, a gap between ZPL and PSB (see inset in fig. \ref{Fig:TEMGridOverview}(b)) indicates mechanical decoupling from low-energy phonon modes. Fig. \ref{Fig:TEMGridOverview}(d) shows distinct off-resonant absorption and emission dipoles. Since the dipoles differ by $87\si{\degree}$, we expect a second excited state to be involved in off-resonant excitation \cite{Jungwirth2016}.  

Next, we use an AFM to investigate the flake topology and its location on the grid. Previous studies suggest that mechanically decoupled quantum emitters in hBN-flakes predominantly emit along the hBN-layers, since their dipole is distorted out-of-plane \cite{Hoese2020}. Therefore, in order to achieve symmetric emission in the reflection and transmission channel we position the flakes inside the hole with the hBN layers parallel to the optical axis. The AFM image depicted in fig. \ref{Fig:TEMGridOverview}(e) discloses the hBN-flake hosting the studied quantum emitter positioned inside the hole of the carbon foil strongly tilted inside the hole. However, the count rate in transmission is still lower compared to the reflection channel. The biased transmission and reflection signal does not necessarily introduce any memory or long-term correlations and true QRNG can still be obtained but rates will be reduced accordingly. The losses in the transmission signal originate mostly from the transmission losses of the grid at the position of the hBN flake. When overlapping microscope and confocal images (see fig. \ref{Fig:TEMGridOverview}(f) and (g)), we find that the hBN flake lies close to the supporting gold grid. Analyzing the transmission properties of empty holes nearby, yield attenuation of the transmission signal at these positions (see fig. \ref{Fig:TEMGridOverview}(h)). Holes further away from the gold support grid show better transmission rates and will be used in future experiments to achieve higher rates. 

\section{Quantum random number generation}

\begin{figure*}[]
\includegraphics[scale=1.]{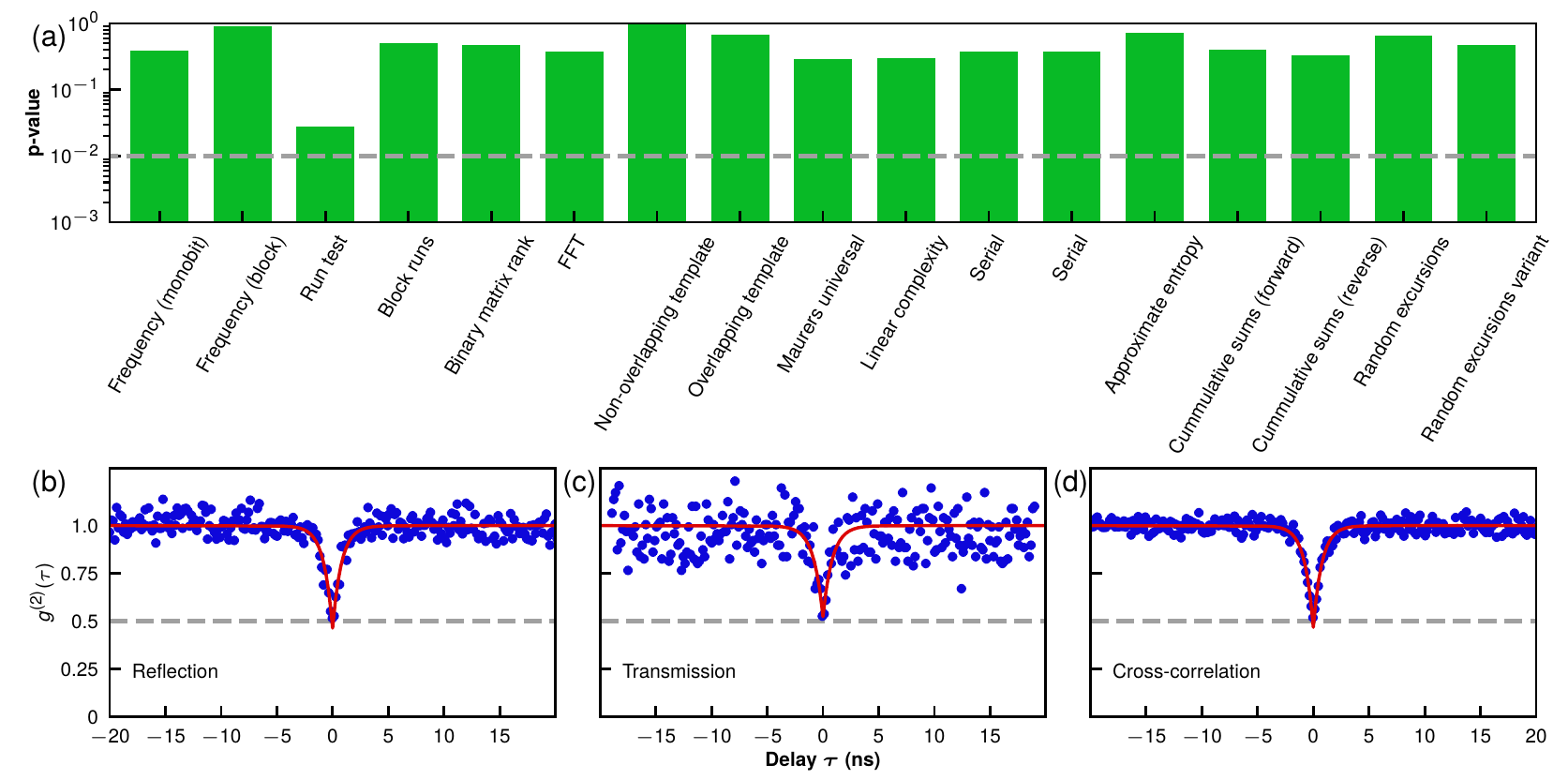}
\caption{\label{Fig:g2_Cross_Trans}QRNG tests. (a) The diagram shows the result of the NIST randomness test of the unbiased bit sequence. The dashed grey line marks the critical p-value above which the sequence can be called random.
(b-d) The results of standard HBT measurements are displayed for the reflection (b) and transmission (c) channel. The right panel (d) shows the result when correlating one APD in reflection with one in transmission. We fit the second-order correlation function to all three measurements (red line).}
\end{figure*}

Finally, we prove the performance as QRNG device by probing the photonic quantum state.  Therefore, we benchmark the randomness of the binary data stream with the NIST test suite \cite{Bassham2010, Ang2020}. To this end, we split the photon stream in reflection into two channels. Each detected photon creates a bit valued 0 or 1 depending on its detection channel. We postprocess the final raw bit sequence to generate independent and equiprobable output bits \cite{Elias1972}. First, we truncate two-bit blocks 11 to 1 and 10 to 0 while discarding the other blocks. Afterwards, we use the von Neumann mapping \cite{vonNeumann1951}, which truncates blocks of 01 to 0 and 10 to 1. This unbiased bit sequence passes all randomness tests of the NIST test suite (see fig. \ref{Fig:g2_Cross_Trans}(a)), thus providing a sequence of true random bits (see supplementary material for details and additional tests \footnote{see supplementary material.}). However, the bit rate is reduced to 21kbit/s compared to 264kbit/s for the raw bit sequence. The result supports our claim that the spontaneous emission of the two SPEs are statistically independent and lead to QRNG. 

Next, we test the second-order cross correlations of the emission in a HBT setup in the reflection and transmission channel, respectively. Both measurements reveal a clear dip at zero time delay (see figs. \ref{Fig:g2_Cross_Trans}(b) and (c)). We fit the data with the second-order correlation function
\begin{equation}
\label{eq:antibunching}
g^{\left(2\right)}\left(\tau\right) = 1- a\exp\left(-\frac{|\tau|}{\tau_0}\right),
\end{equation}
with the time delay of the photon arrival times $\tau$, lifetime $\tau_0$ and scaling factor $a$. Both fits yield values at zero time delay of $g^{\left(2\right)}\left(0\right) = 0.46 \pm 0.03$ and $g^{\left(2\right)}\left(0\right) = 0.52 \pm 0.12$ and lifetimes of $\tau_0 = 0.79 \pm 0.07 \si{\nano\second}$ and $\tau_0 = 0.8 \pm 0.3 \si{\nano\second}$. Note, that the correlation measurement in transmission is noisier due to reduced count rate.  

Afterwards, we examine cross correlations between reflection and transmission channel with one APD in each channel. The defect center emits only one photon at a time in a random direction. Hence, spontaneously emitted photons enter the reflection or transmission channel randomly distributed with equal probability. So, there can either be one photon in the reflection channel and none in the transmission channel, or vice versa. We denote these two different states as $\ket{1}_\text{R}\ket{0}_\text{T}$ and $\ket{0}_\text{R}\ket{1}_\text{T}$. An emitted photon is in a superposition of these states,
\begin{equation}
\label{eq:state}
\ket{\psi} = \frac{1}{\sqrt{2}}\left( i\ket{1}_\text{R}\ket{0}_\text{T} + \ket{0}_\text{R}\ket{1}_\text{T} \right),
\end{equation}
before it is detected by an APD in one of the two channels. Each detected photon can be interpreted as a single random bit valued 0 or 1 according to whether it is detected in reflection or transmission, analogous to beamsplitter-based QRNG \cite{Jennewein2000}. We note that the state of a single photon before being detected in one of both channels equals the state of a photon after passing a beamsplitter in a HBT experiment \cite{Oberreiter2015}. When we now introduce an imbalance between reflection and transmission channel, the state of equation (\ref{eq:state}) changes to 
\begin{equation}
\label{eq:imb-state}
\ket{\psi} = i\sqrt{R}\ket{1}_\text{R}\ket{0}_\text{T} + \sqrt{T}\ket{0}_\text{R}\ket{1}_\text{T},
\end{equation}
where R and T denote the relative photon number distribution in the reflection and transmission channel. The single-photon count rate distribution is divided by 9\% in channel T and 91\% in channel R which limits the overall bit rate but not its randomness nor the second-order correlations, in analogy to imbalanced beamsplitters in HBT setups \cite{Loudon2000}. To compensate inequalitites in the count rates further improvement on the dipole alignment and on the transmission losses could be obtained. Figure \ref{Fig:g2_Cross_Trans}(d) shows the characteristic antibunching at short time delay for cross correlations between the transmission and detection channel. Fitting the correlation function with equation (\ref{eq:antibunching}) yields $g^{\left(2\right)}\left(0\right) = 0.47 \pm 0.02$ and an excited state lifetime $\tau_0 = 0.77 \pm 0.04 \si{\nano\second}$, in agreement with results from HBT experiment in reflection and transmission. Please note, that stimulated emission processes would fundamentally limit the performance, but can be neglected due to the off-resonant excitation scheme used.

\section{Discussion and Outlook}

In conclusion, we showcase the creation of single photon randomness with quantum emitters in hBN where the randomness arises from the symmetry of the dipole emission and the random character of the spontaneous emission process based on the principles of quantum mechanics. By exploiting the random single photon emission directionality, we split the emission into a reflection and a transmission channel by preparing the hBN-flake on a grid situated between two high-NA objectives for photon collection. Each channel is designated as 0 or 1 in order to create a binary random number code. Our realization of QRNG with SPEs does not rely on any beam-splitting optics. Therefore, the rates could be increased by a factor of two as compared to architectures relying on beam splitters. Based on the extracted lifetime of about 0.8 ns and assuming that almost all emitted photons are collected with one of the two channels puts raw bit rates beyond gigabit per second into reach. Furthermore, we operate in a regime of high entropy per raw bit, higher than achievable with classical light sources, where the detectors are not saturated nor limited by dark times. We show that photons emitted from two independent SPEs withstand the tests for randomness. The underlying single photon characteristic of each SPE could be used as additional security test against eavesdropping \cite{Chen2019}. 

We also show that the SPEs optical properties persist against nanomanipulation of the hBN-host. The nanomanipulation enables further optimization of the absorption and emission polarization with respect to the experimental setup. The combination with pick-and-place techniques \cite{Schell2017, Fehler2020} opens the door for hybrid integration into photonics \cite{Kim2020} accompanied by further miniaturization and improved bandwidth \cite{Vogl2019, Froech2020, Haeussler2020}. The studied defect centers show the signatures of mechanical decoupling from low-energy acoustic phonons, as indicated by an energy gap between the ZPL and the PSB. In the future, the resulting narrow spectral linewidth of these emitters, known to be within the FTL at room temperature under resonant excitation, could be used to marry QRNG with photonic quantum technologies \cite{Caldwell2019, Lee2020, Holmes2020} and coherent control \cite{Konthasinghe2019}.

\medskip


\section*{Acknowledgments}
M.K.K. and A.K. acknowledge support of the IQst. A.K. is grateful for support from the H2020 Marie Curie ITN project LasIonDef (GAn.956387). A.K. acknowledges support of the European Regional Development Fund (EFRE) program Baden Württemberg. M.H. acknowledges support from the Studienstiftung des deutschen Volkes. The AFM was funded by the DFG. We thank Prof. Kay Gottschalk and Frederike Erb for their support.

\section*{Author contributions}
M.H., M.K.K. and A.K. conceived the experiments. All measurements were performed and analyzed by M.H., M.K.K. and F.B. with help from N.L. and K.G.F. for AFM characterization. All authors discussed the results. M.H., M.K.K. and A.K. wrote the manuscript, which was discussed and edited by all authors.


\bibliography{hBN-Transmission}

\end{document}